\documentclass[%
 reprint,
superscriptaddress,
 amsmath,
 amssymb,
 aps,
 prl,
]{revtex4-2}

\usepackage{graphicx}
\graphicspath{{Final_plots/Fig1/}{Final_plots/Fig2/}{Final_plots/Fig3/}{Final_plots/Fig4/}{Final_plots/Appendix}}
\usepackage{comment}
\usepackage{dcolumn}
\usepackage{bm}
\usepackage{hyperref}

\usepackage{xfrac}
\usepackage{siunitx}
\usepackage[T1]{fontenc} 
\usepackage[dvipsnames]{xcolor}
\usepackage{bbm}
\colorlet{ForestGreen}{Black}
\begin{document}

\preprint{APS/123-QED}

\title{Readout of a solid state spin ensemble at the projection noise limit}

\author{Rouven Maier}
 \affiliation{3rd Institute of Physics, University of Stuttgart, 70569 Stuttgart, Germany.}  
 \affiliation{Max Planck Institute for Solid State Research, 70569 Stuttgart, Germany.}

\author{Cheng-I Ho}
 \affiliation{3rd Institute of Physics, University of Stuttgart, 70569 Stuttgart, Germany.}  
 \affiliation{Center for Integrated Quantum Science and Technology (IQST), 70569 Stuttgart, Germany.}


\author{Andrej Denisenko}
 \affiliation{3rd Institute of Physics, University of Stuttgart, 70569 Stuttgart, Germany.}

\author{Marina Davydova}
 \affiliation{Fraunhofer Institute for Applied Solid State Physics (IAF), 79108 Freiburg, Germany}

\author{Peter Knittel}
 \affiliation{Fraunhofer Institute for Applied Solid State Physics (IAF), 79108 Freiburg, Germany}


 
\author{J\"org Wrachtrup}
 \affiliation{3rd Institute of Physics, University of Stuttgart, 70569 Stuttgart, Germany.}
 \affiliation{Max Planck Institute for Solid State Research, 70569 Stuttgart, Germany.}
 \affiliation{Center for Integrated Quantum Science and Technology (IQST), 70569 Stuttgart, Germany.}
 
 \author{Vadim Vorobyov}
 \email[]{v.vorobyov@pi3.uni-stuttgart.de}
 \affiliation{3rd Institute of Physics, University of Stuttgart, 70569 Stuttgart, Germany.}  
 \affiliation{Center for Integrated Quantum Science and Technology (IQST), 70569 Stuttgart, Germany.}

\begin{abstract}
Spin ensembles are central to quantum science, from frequency standards and fundamental physics searches to magnetic resonance spectroscopy and quantum sensing. 
Their performance is ultimately constrained by spin projection noise, yet solid-state implementations have so far been limited by much larger photon shot noise. 
Here we demonstrate a direct, quantum non-demolition readout of a mesoscopic ensemble of nitrogen-vacancy (NV) centers in diamond that surpasses the photon shot-noise limit and approaches the intrinsic spin projection noise. 
By stabilizing the $^{14}$N nuclear spin bath at high magnetic fields and employing repetitive nuclear-assisted spin readout, we achieve a noise reduction of $\SI{3.8}{dB}$ below the thermal projection noise level. 
This enables direct access to the intrinsic fluctuations of the spin ensemble, allowing us to directly observe the signatures of correlated spin states. 
Our results establish projection noise-limited readout as a practical tool for solid-state quantum sensors, opening pathways to quantum-enhanced metrology, direct detection of many-body correlations, and the implementation of spin squeezing in mesoscopic solid-state ensembles.
\end{abstract}

\maketitle

\begin{figure}[t]
    \includegraphics{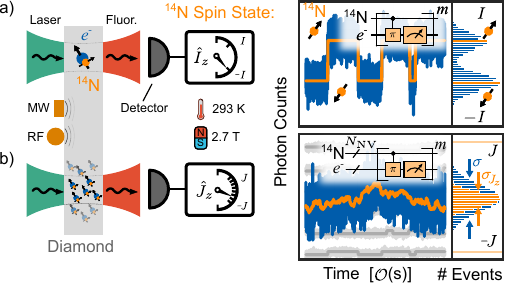}
    \caption{\label{fig:schematic} \textbf{Modeled Quantum state readout: From single NVs to NV ensembles} 
    \textbf{a)} A single Nitrogen-vacancy (NV) center in a confocal microscope at a magnetic field of $B_0 = \SI{2.7}{T}$ is controlled by laser, microwave (MW) and radio frequency (RF) control and the fluorescence is read out by the detector.
    The Nitrogen spin $I_z$ undergoes stochastic spin fluctuations $| \uparrow \rangle \leftrightarrow |\downarrow \rangle$, producing a random telegraph signal (orange line). 
    Photon shot noise (blue line) is introduced by the optical readout. 
    \textbf{b)} In the signal of an ensemble of NV centers, the individual telegraph like signals (gray lines) are summed to a net magnetization $\langle J_z \rangle $ (orange line), where individual spin states can no longer be determined.
    Typically, the fluctuations $\sigma_{J_z}$ (projection noise) are hidden in the photon shot noise (blue line).
    }
\end{figure}

\begin{figure*}
    \includegraphics[]{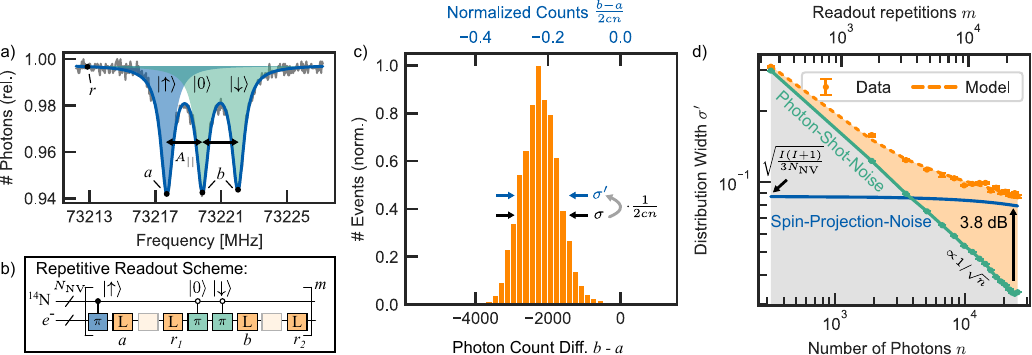}
    \caption{\label{fig:N_ssr} \textbf{Surpassing the photon shot noise.}
    \textbf{a)} Optically detected magnetic resonance (ODMR) spectrum of the NV center. 
    The nuclear spin selective electron transitions are separated by the hyperfine coupling $A_\mathrm{||} = \SI{2.16}{MHz}$.
    \textbf{b)} Pulse sequence of the repetitive readout of $N_\mathrm{NV}$ nuclear spins.
    The nuclear spin states are transferred onto the electron spin via selective $\pi$-rotations, before a readout laser pulse is applied.
    The detected photons ($b - a$) are summed up over $m$ repetitions of the readout.
    Reference measurements ($r$) are added in between the spin readouts.
    The respective photon counts $a, b$ and $r$ are visualized in panel a).
    \textbf{c)} Typical photon histogram of 3000 experiments obtained by $b - a$.
    The (normalized) width ($\sigma^\prime$) $\sigma$ is caused by the photon shot noise and the spin projection noise. 
    \textbf{d)} An increase of photon counts $n$ during the readout reduces the photon shot noise (green) as $1/\sqrt{n}$.
    For high photon counts, a transition from the photon shot noise dominated regime to the projection noise (blue) dominated regime is observed, where the photon shot noise is $\SI{3.76(1)}{dB}$ lower than the observed projection noise.
    The projection noise is given by the statistical polarization $\sigma_{\tilde{J}_z} = \sqrt{\frac{I(I+1)}{3N_\mathrm{NV}}}$, which reduces during the readout due to $T_1$ relaxation of the nitrogen spins.
    }
\end{figure*}

\section{Introduction}
Ensembles of quantum spins represent a foundational platform in quantum physics with a wide range of applications, ranging from new frontiers in frequency standards \cite{beeks2021thorium,higgins2025temperature, zhang_frequency_2024}, inertial navigation \cite{kanegsberg1978nuclear}, searches for dark matter \cite{sushkovDark2023}, to widespread nuclear magnetic resonance (NMR) spectroscopy \cite{glennCASR2018, bucher_quantum_2019} and magnetic resonance imaging (MRI) \cite{bruckmaier_imaging_2023, briegel2025optical} applications. 
Atomic spin ensembles \cite{budker2007optical, patton2014all, behbood2013real} are the gold standard with regard to magnetic field sensitivity due to sub-projection noise limited readout \cite{koschorreck2010sub, hostenmeasurement2016, janviercompact2022, zhousecondscale2020} and long coherence times, allowing for example the detection of human brain activity \textit{in-vivo} \cite{boto2018moving}.
Following an early theoretical proposal \cite{Wineland:1994aa}, this efficient readout in combination with a high degree of spin control paved the way to even overcome the projection noise limit (standard quantum limit) by applying spin squeezing \cite{PhysRevLett.109.253605, robinsondirect2024, louchet-chauvetentanglementassisted2010}.
Extending these capabilities to solid-state spin ensembles enables scalability and integrable quantum sensors \cite{garsi2024three, smits2019two, soshenko2021nuclear} with applications ranging from nanoscale magnetic resonance spectroscopy, bioimaging \cite{allertAdvances2022, loretzNanoscale2014, staudacherNMR2013, rugarProtonNMR2015, lovchinsky2016nuclear} and nanoscale scanning probe \cite{maletinsky2012robust, qiu_nanoscale_2022} to studies of correlated sensors \cite{rovny2022nanoscale, ji_correlated_2024,lei_quantum_2025}.
Among solid-state platforms, nitrogen-vacancy (NV) centers in diamond combine long coherence times, room-temperature operation and optical addressability \cite{doherty_nitrogen-vacancy_2013}.
Yet, unlike their atomic counterparts and despite significant progress in solid state ensembles, including NVs, readout schemes have so far remained constrained by photon shot noise \cite{zhang2021diamond, taylor2008high, arunkumar2023quantum, ebel2021dispersive}, obscuring intrinsic quantum fluctuations. 
This restriction not only limits the sensor performance, but also precludes direct observation of collective quantum effects such as correlated spin noise or spin squeezing \cite{wu2025spin, martin2023controlling, gao2025signal}.
Here, we overcome this barrier by realizing a projection noise limited readout in a solid state spin ensemble.
Utilizing repetitive nuclear-assisted measurements, formerly developed for single spins \cite{Neumann.2010, maurer2012room}, and applying it to a mesoscopic spin ensemble of NV centers at high magnetic field, we achieve a noise reduction of \SI{3.8}{dB} below the photon shot noise level.
We probe the crossover between photon shot noise and spin projection noise limited regimes in the optical readout of the unpolarized ensemble and show the traces of correlated states in the variance of the readout noise distribution and propose it for correlated quantum sensing using sub-micron-scale ensembles of NV centers.
The sensitivity improvement up to several orders of magnitude for quantum sensing protocols through projection noise limited readout is discussed and the optimum readout conditions are identified.
By closing the gap between atomic and solid-state ensemble measurements, our work establishes projection noise limited readout as a new resource for solid-state quantum sensing and lays the foundation for quantum-enhanced metrology, correlated sensing and many-body studies in diamond spin ensembles.

\section*{Results}
All experiments are carried out on a home-built confocal microscope at room temperature (Fig. \ref{fig:schematic}).
The (111) cut diamond chip containing as grown preferentially aligned thin ensembles of $^{14}$NV centers is placed in a superconducting magnet, with a bias magnetic field of $B_0 = \SI{2.7}{T}$ aligned to the principal NV axis.
Microwave (MW) control of the electron spin at \SI{73}{GHz} is achieved using a $\lambda/2$ bow-tie resonator, as described in \cite{Maier.17032025}.
Nuclear spins are controlled via radio frequency (RF) from a \SI{50}{\mu m} copper wire spanned across the sample.
Emitted photons are detected using an avalanche photodiode (APD).
See \cite{SI} for further details.

\begin{figure*}[htp]
    \includegraphics[]{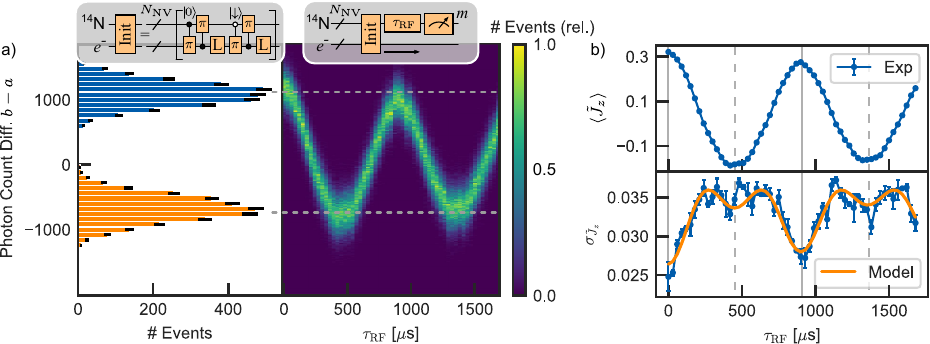}
    \caption{\label{fig:nitrogen} \textbf{Nitrogen spin control.} 
    \textbf{a)} The ensemble of nitrogen spin states can be polarized by swapping the polarized electron spin state onto the nitrogen spins with a series of controlled $\pi$ rotations.
    Rabi oscillations of the nitrogen spins are driven by resonant RF.
    The histograms of the detected photons $b-a$ shift accordingly.
    Full inversion is not achieved due to a limited initialization fidelity of the electron spin.
    \textbf{b)} The extracted spin distribution width $\sigma^\prime$ follows the oscillations of the spin projection noise $\sigma^2_{\tilde{J}_z} \propto p(1-p)$, where $p$ is the readout probability into the nitrogen eigenstates.
    Full recovery of the polarized $\sigma_{\tilde{J}_z}$ after a full inversion (dashed line) is not observed due to faster spin decay in the nitrogen $|0\rangle$ state during the readout (see main text).
    }
\end{figure*}

The average collective nuclear spin $\tilde{J}_z = \frac{J_z}{N_\mathrm{NV}} = \frac{1}{N_{\mathrm{NV}}}\sum_{i=1}^{N_\mathrm{NV}} I_z^{(i)}$ of the $N_\mathrm{NV}$ intrinsic nitrogen spins with spin $I = 1$ of an ensemble of NV centers is measured by optical quantum-non-demolition measurements, where $I_z^{(i)}$ is the spin projection along the $z$ axis of the individual nitrogen spin $i$.
Spin state readout of the nitrogen hyperfine levels is achieved by repeated mapping onto the reinitialized electron spin states $S_z$ of the NV centers via a spin-selective narrowband MW $\pi$ pulse, which is then read out and reinitialized optically (Fig. \ref{fig:N_ssr} a,b). 
This way, the corresponding photon numbers $a$ and $b$ are obtained, providing information about the population of the three nitrogen spin levels $\{ |\uparrow \rangle \}$ and $\{|0\rangle, |\downarrow \rangle \}$.
To increase the signal contrast and remove influence of slow drifts of laser power, the main signal is obtained by subtracting the counts $b -a$.
As both, the $|0\rangle$ and $|\downarrow \rangle$ states are read out together, the three nitrogen states are mapped onto a pseudo spin-1/2 system with the eigenstates $\{|\uparrow\rangle, |\downarrow \rangle \}$. 
The total readout contrast for each measurement is calculated via $c = 2-\frac{a+b}{n}$, where $n$ is the baseline photon count $(r_1 + r_2)/2$.
The nuclear spin state is left almost undisturbed by this process, allowing many repetitions $m$ of the readout to accumulate fluorescence signal of the photon count difference $b - a$.
This yields a Skellam distribution with photon shot noise limited standard deviation $\sigma_n = \sqrt{1-\frac{c}{2}}\sqrt{2n}$ (Fig. \ref{fig:N_ssr}c).

This distribution enables the reconstruction of the nuclear spin distribution properties as described below.
The average $\langle \tilde{J}_z \rangle$ is given simply by a linear expression $\langle \tilde{J}_z \rangle = k_0 \langle b - a \rangle $, where $k_0 = 1/2 n c$, where $c$ is the optical contrast of the readout between states $|\downarrow\rangle^{\otimes N}$ and $|\uparrow \rangle^{\otimes N}$.
The influence of the spin projection noise $\sigma_{\tilde{J}_z}$ and photon shot noise $\sigma_n$ result in a variance sum for the observed signal distribution:
\begin{align}
    \sigma^2 &= \sigma_n^2 + (2 n c \sigma_{\tilde{J}_z})^2 \thickspace .\label{eq:sigma}
\end{align}
We analyze the standard deviation of the obtained signal multiplied by the scaling factor $k_0$: $\sigma^\prime = \sigma/2 n c$. 
By gradually increasing the number of detected photons $n$ through an increase of the number of repetitive readouts $m$ from 1250 to 25000 we probe the crossover in the readout of the spin ensemble from photon shot noise to spin projected noise limited regime depicted on Fig. \ref{fig:N_ssr}d).
In the limit of small collected number of photons, the normalized standard deviation is dominated by the photon shot noise and scales as $\sigma^\prime \rightarrow 1/\sqrt{n}$.
By increasing the number of readout repetitions, $\sigma^\prime$ begins to deviate from the photon noise scaling and levels with uncertainty of the thermal spin distribution (spin projection noise) given by $\sigma^\prime \rightarrow \sqrt{\frac{I(I+1)}{3 N_\mathrm{NV}}}$ (See \cite{SI} for details). 
In this regime the direct readout of the ensemble spin state statistics becomes possible.

The maximum duration of the readout is limited by the longitudinal relaxation time $T_1$ of the probed spins.
As it increases quadratically with the applied magnetic field \cite{Neumann.2010}, we apply a magnetic field of $B_0 = \SI{2.7}{T}$ which enables the realization of more than $\sim \SI{5e4}{}$ repetitive readouts required to reach the projection noise limit.
To account for a residual spin relaxation during the readout (Fig. \ref{fig:N_ssr} d) we adapt a model, which captures the reduction in spin projection noise under slightly perturbing measurements. 
First, we estimate the correlation function of the spin ensemble under measurements $C(\tau) = \langle \tilde{J}_z(t) \tilde{J}_z(t+\tau)\rangle_t = \sigma_{0}^2 e^{-|\tau|/T_1}$ as an exponential decay with characteristic time $T_1$, confirmed in a separate measurement (see \cite{SI}). 
Second, the standard deviation of the spin distribution under perturbing measurement is estimated as a standard deviation of a stochastic process averaged over measurement time $T\sim T_1$ comparable to the relaxation time (see \cite{papoulis2002probability} and \cite{SI} for more details).
The projection noise $\sigma_{\tilde{J}_z}$ in the measured data can be fitted as a function of the nuclear spin relaxation time $T_1$ under readout and the number of nitrogen spins $N_\mathrm{NV}$ as:
\begin{align}
    \sigma_{\tilde{J}_z} = \sqrt{\frac{1}{3}} \sqrt{\frac{I(I+1)}{3N_\mathrm{NV}}} \sqrt{\frac{2T_1^2}{T^2} \left(\frac{2T}{T_1} + e^{-T/T_1} -1\right)}
\end{align}
The factor of $\sqrt{1/3}$ is introduced as a side effect of reading out the population of the three nitrogen spin states by mapping them onto only two electron spin states \cite{SI}.
Finally, we take into the account the effect of the avalanche photodiode (APD) saturation at high photon detection rates leading to an artificial decrease of the width of the detected photon distribution.
This is accounted for, by adding a brightness-dependent verified correction factor $k$ to the model ($k \sim 0.89 - 0.99$), yielding:
\begin{align}
        \sigma^\prime &= k \sqrt{\left( \sigma^\prime_n \right)^2 + \sigma_{\tilde{J}_z}^2} \thickspace.
\end{align}


\begin{figure*}[t]
    \includegraphics{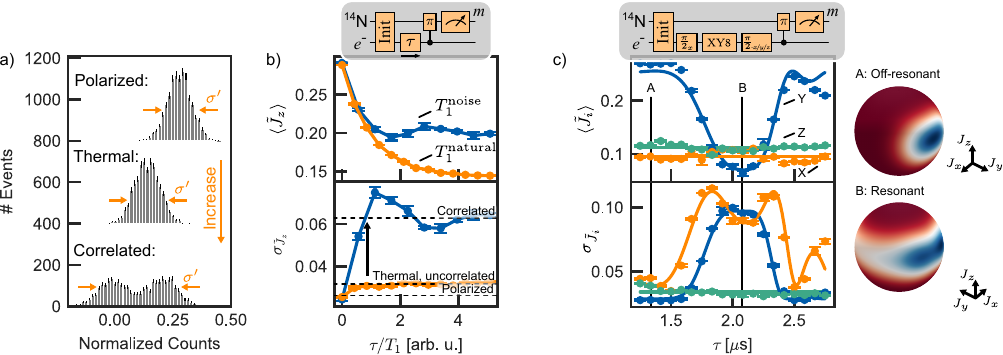}
    \caption{\label{fig:T1} \textbf{Detection of correlated noise sources.}
    \textbf{a)} Spin distributions of different prepared spin states. 
    The spin dominated distribution width $\sigma^\prime$ increases from polarized to thermal to spatially correlated state.
    \textbf{b)} Comparison of spin decay caused by natural $T_1$ processes (orange) and a noisy MW drive (blue).
    On average, the electron spin $\langle \tilde{J}_z \rangle$ is decohered into a steady state (upper panel).
    The projection noise $\sigma_{\tilde{J}_z}$ after the natural $T_1$ process increases to the thermal, spatially uncorrelated state, while spatial correlations are observed for a noisy common drive (lower panel).
    The timescale is normalized by the effective decay time.
    \textbf{c)} Detection of an applied oscillating RF field with frequency $f = \SI{250}{kHz}$ via noise spectroscopy.
    By choosing the axis of the final $\pi/2$ pulse a quantum tomography of the spin state along X (orange), Y (blue) and Z (green) after interaction with the external field is achieved.
    When coupling to the external field (at point B), the polarization along Y is destroyed, while X and Z remain unpolarized (upper panel).
    The correlated spin distribution is observable and shows an increase in $\sigma_{\tilde{J}_x}$ and $\sigma_{\tilde{J}_y}$ as the spins are redistributed in the equatorial plane through the interaction with the oscillating signal (lower panel). 
    The Husimi $Q$-function of the reconstructed spin distribution is visualized on a sphere, showing the delocalization of the spin states in the equatorial plane.
    For details about the fitting model (solid lines) and the spin state reconstruction, see \cite{SI}.}
\end{figure*}

As a result, this method enables the direct determination of active NV centers in the confocal spot of the laser from the model fit parameters ($N_\mathrm{NV} = \SI{31(3)}{}$).
We probe it in three various NV density spots on our sample, yielding a clear linear correlation with observed fluorescence levels (see \cite{SI}).
It also should be noted, that as the relative photon shot noise and the thermal projection noise both decrease as $\sim 1/\sqrt{N_\mathrm{NV}}$, this method of projection noise limited readout is generally applicable to arbitrary ensemble sizes.

With the established readout we probe the coherent control of the spin ensembles at hand. 
First, the spin ensemble is initialized into $|\uparrow\rangle$ via repeatedly swapping the polarization from the electron spin to the nitrogen spin through the application of selective $\pi$ pulses (Fig. \ref{fig:nitrogen} a).
Applying resonant radio frequency (RF) pulses between $|1\rangle$ and $|0\rangle$ allows direct spin control of the nitrogen spin, as long-lived Rabi oscillations are observed (Fig. \ref{fig:nitrogen} a).
A deviation from $\langle \tilde{J}_z \rangle = 0.5$ after the initialization is due to an initialization infidelity of the nitrogen spin states, with some remaining population in $|0\rangle$ and $|-1\rangle$.
Additionally, limited initialization fidelity of the electron spin prevents a full population inversion, as the RF is only resonant to the nitrogen transition, if the electron is initialized into the correct charge and spin state (i.e. NV$^-$; $m_s = 0$).
The other fraction of nitrogen spins stays polarized during the experiment.
Due to repetitive initialization of the electron spin during the readout step, we still read out the spin state of all nitrogen spins.
As a result of the projection noise limited readout, we are able to track the spin projection noise $\sigma_{\tilde{J}_z}$ next to the average spin state $\langle \tilde{J}_z \rangle$ during the drive of the nuclear spin oscillations (Fig. \ref{fig:nitrogen} b).
The projection noise $\sigma_{\tilde{J}_z}$ is directly calculated according to Eq.\ref{eq:sigma}.
The spin projection noise follows the expected projection noise relation of $\sigma_{\tilde{J}_z} \sim \sqrt{p(1-p) N_{NV}}$, where $p$ is the probability to measure the eigenstates.
Starting with a fully polarized state, the width of the distribution is minimized, limited by a residual decay of the spin state during the readout and the infidelity of the polarization. 
During the Rabi oscillations, the width reaches its maximum when the ensemble state crosses the equator plane, before it is reduced again, following the partial inversion of the population. 
The limited dip of $\sigma_{\tilde{J}_z}$ at the inversion is due to an increased polarization decay $T_1$ of $|0\rangle$ compared to $|1\rangle$ during the readout \cite{SI}.

Next we employ the readout to explore novel sensing capabilities of the electron spin ensemble on spatially correlated sensing.
The initialized state of the nitrogen is used to partially swap the sensing state of the electron spin there to achieve nuclear spin assisted electron spin readout and apply it to two typical sensing protocols: $T_1$ relaxometry \cite{mzyk_relaxometry_2022,han_solid-state_2025} and oscillating field quantum sensing \cite{alvarez_measuring_2011,de_lange_single-spin_2011} (Fig \ref{fig:T1}).

In the case of $T_1$ relaxometry we compare spatially correlated noise created by a wideband microwave signal and uncorrelated intrinsic phonon induced noise. 
For a correlated noise source, a microwave signal generated as a sum of ten fully randomized frequency sources in a 3 MHz band around the NV center electron spin resonance transition frequency is applied, thus inducing fast electron spin decay. 
For the uncorrelated noise source we rely on the intrinsic NV center phonon induced translational relaxation of spin sublevels, as they occur independently and incoherently for each NV center in the ensemble.
The observed spin decay and its standard deviation in the timescale normalized by the respective total relaxation time $T_1$ are plotted in Fig. \ref{fig:T1} a) and b). 
While the average spin state $\langle \tilde{J}_z \rangle$ decays similarly, a distinction in the obtained $\sigma_{\tilde{J}_z}$ is seen, indicating a clear sign of correlations induced by the correlated noise source.
The decay curve of the noise induced experiment follows the expected Bessel-like average of a randomly driven spin.
The thermal spin state induced by uncorrelated noise is significantly narrower compared to the state induced by the correlated noise.
This shows, that having access to the direct spin readout can allow the investigation of spatial correlations in environmental effects using sub-micronscale ensembles of sensors in microscopy applications.

In a second typical scenario, $T_2$ relaxometry is used to detect correlated oscillating noise generated by a monochromatic source with a stochastic phase $S = B_\mathrm{osc} \cos (2\pi f t + \lambda)$, where frequency $f = \SI{250}{kHz}$ and amplitude $B_\mathrm{osc} = \SI{1.84(7)}{\mu T}$ at the position of the NV center. 
After preparing the initial electron spin state $\rho_S =(\mathbf{1_e} + \sigma_y/2)^{\otimes N}$ a XY8 dynamical decoupling sequence is applied and the interaction with the noise is controlled by adjusting the inter-pulse spacing $\tau$ with resonance at $\tau = 1/(2f)$ \cite{degen_quantum_2017}.
The marginal spin projection distributions $\tilde{J}_x,\tilde{J}_y, \tilde{J}_z$ after the interaction with the noise are obtained by applying different pulses \{$\pi/2_y$, $\pi/2_{-x}$, $\varnothing$\} followed by $\langle J_z\rangle$ readout, as shown in Fig. \ref{fig:T1} c). 
Apart from the usually detected average value $\langle \tilde{J}_i \rangle$ the correlated spin fluctuations along the X, Y and Z axis induced by the correlated oscillating noise are probed.
Again, direct access to the spin distributions by projection noise limited readout allows for a more sophisticated study of the effects of the noisy environment on the sensor spins, by clearly showing delocalization in the equatorial plane.  
To better visualize the ensemble spin state we plot the Husimi $Q$-function of the regularized reconstructed spin distribution of the active spins on a sphere. 
For details regarding the reconstruction method, see \cite{SI}.


\begin{figure}[t]
    \includegraphics{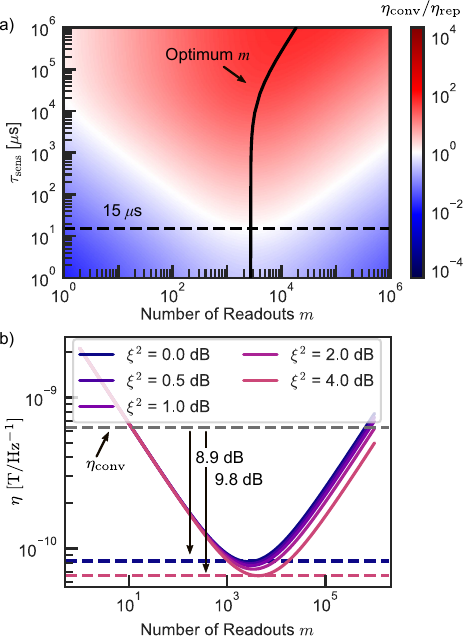}
    \caption{\label{fig:sensitivity} \textbf{Calculated sensitivity improvement through projection noise limited readout.} 
    \textbf{a)} For sensing protocols with $\tau_\mathrm{sens} > \SI{15}{\mu s}$ the repetitive readout scheme improves sensitivity of a superposition (polarized) state readout (red area) compared to the conventional, single readout.
    For $\tau_\mathrm{sens}< 10^4 \mu \mathrm{s}$ the optimum repetition number $m$ is at 2700 readouts.
    For larger $m$, the spin projection noise is reached, limiting signal improvement, while the protocol duration is increased, leading to a declining sensitivity.
    \textbf{b)} Spin squeezing $\xi^2$ can further improve the sensitivity by reducing the spin projection noise limit.
    Here, the absolute values for $N_\mathrm{NV} = 100$ and $\tau_\mathrm{sens} = \SI{1}{ms}$ are shown.
    }
\end{figure}

After establishing a readout technique that can reach the spin projection noise level, here we investigate its applicability to common quantum sensing protocols with sensing time $\tau_\mathrm{sens}$, especially with respect to the sensitivity.
An improved sensitivity is characterized by a lower value of $\eta$, which is given by
\begin{align}
    \eta \propto \frac{1}{c_\mathrm{eff} n \sqrt{\tau_\mathrm{sens}}}\sqrt{\frac{\tau_\mathrm{sens} + \tau_\mathrm{other}}{\tau_\mathrm{sens}}} \sigma \thickspace, 
\end{align}
where the effective contrast $c_\mathrm{eff}$ and experimental overhead $\tau_\mathrm{other}$ (e.g. initialization and readout) are counteracting the positive effect of a reduced noise level $\sigma$. 
Based on the experimental parameters extracted in this work\cite{SI}, the sensitivity of sensing protocols with a conventional readout consisting of a single direct electron spin readout with a laser pulse is compared to the sensitivity of the nuclear spin assisted repetitive readout.
As example, a sensing protocol in the linear regime around the sensing state $\rho_S =(\mathbf{1_e} + \sigma_y/2)^{\otimes N}$ (i.e. maximum spin projection noise) is investigated.
The ratio $\eta_\mathrm{conv}/\eta_\mathrm{rep}$ as a function of the number readouts $m$ and the duration of the sensing protocol $\tau_\mathrm{sens}$ is shown in Fig. \ref{fig:sensitivity} a).
Already for sensing protocols with $\tau_\mathrm{sens} > \SI{15}{\mu s}$ (typical for $T_2$ based sensing protocols) the repetitive readout can improve the sensitivity, while for very long measurement sequences the improvement can reach several orders of magnitude.
For a given $\tau_\mathrm{sens}$, $\eta_\mathrm{rep}$ first decreases as $\sigma$ approaches the spin projection noise limit.
After this point, $\sigma$ cannot be improved further by increasing $m$, while simultaneously increasing the experimental overhead, leading to an increase of $\eta_\mathrm{rep}$.
One avenue to further increase the sensitivity of spin projection noise limited readouts is the use of spin-squeezing techniques.\cite{wu2025spin}
The influence of spin squeezing up to \SI{4}{dB} on $\eta_\mathrm{rep}$ as a function of the readout repetitions is shown in Fig. \ref{fig:sensitivity} b).
Previously demonstrated values of $\eta^2 = \SI{0.5}{dB}$ do not significantly improve the sensitivity.



\subsection*{Conclusion and Discussion}
In summary, we have realized projection noise-limited readout of a solid-state spin ensemble, surpassing the thermal spin projection noise floor by $\SI{3.8}{dB}$.
This advance provides a long-sought capability: direct access to the fundamental fluctuations of solid state mesoscopic spin systems.

Our approach can improve the sensitivity of NV based quantum sensors up to several orders of magnitude and opens the opportunity to harness the benefits of spin-squeezing in solid state quantum sensing.
This will enable rapid NMR, MRI, magnetometry and relaxometry measurements based on quantum sensors, where large averaging times are limiting the applicability. 
Another exciting avenue is the microscopy of temporally and spatially correlated signals, enabled by projection noise limited readout.
We envision a widefield microscope, where each diffraction-limited pixel is detecting spin distributions, which can directly be analyzed to obtain a correlation map.
Possible areas of application include the investigation of interaction types in magnetic materials, phase transitions or electronic chip failures.
Further, the application of solid state spin ensembles in the field of quantum simulations \cite{davisprobing2023} is directly linked to the readout of the spin states.
A projection noise limited readout for example allows the extraction of 2$^\mathrm{nd}$, 4$^\mathrm{th}$ and higher order cumulants to observe non-trivial, non-classical and non-Gaussian many body spin states \cite{dubostefficient2012}.

\subsection*{Acknowledgments}
We acknowledge funding from the German Federal Ministry of Education and Research (BMBF) through the projects Clusters4Future QSens and DiaQnos, and the German Federal Minsitry of Research, Technology and Space through projects Qsolid and QSi2V.
We further acknowledge funding from the European Union through project C-QuENS (grant agreement no. 101135359), Amadeus (grant agreement no. 101080136) and Spinus (grant agreement no. 101135699), as well as the Carl-Zeiss-Stiftung via QPhoton Innovation Projects and the Center for Integrated Quantum Science and Technology (IQST).
R.M. acknowledges support from the International Max Planck Research School (IMPRS).


\bibliography{references}
\end{document}


\setcitestyle{numbers}
\maketitle

\maketitle

\begin{figure}[ht]
    \centering
    \includegraphics{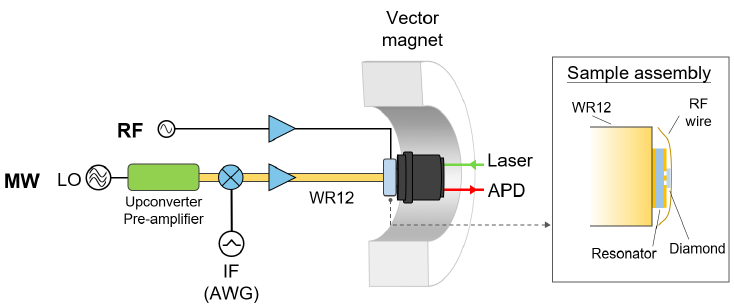}
    \caption{\textbf{Experimental Setup.}
    The experiments are performed on a confocal, room-temperature, ensemble-NV setup. 
    The NV-containing diamond is placed inside a superconducting magnet (2.7 T) and optically addressed with a green laser, focused through a high-NA objective. 
    Red fluorescence is collected and detected using an avalanche photodiode (APD). 
    Microwave (MW) pulses at 73 GHz are generated by upconverting the local oscillator (LO) signal six times and mixing it with pulsed MW from the arbitrary waveform generator (AWG). 
    RF control of the nuclear spins is applied to the diamond via a copper wire on its backside. 
    Inset: detailed view of the diamond and resonator.
    }
    \label{fig:Setup}
\end{figure}

\section*{Supplementary Note 1: Experimental setup}
A schematic of the experimental setup is provided in Supplementary Fig. \ref{fig:Setup}. 
Optical excitation of the NV center is achieved with a pulsed $\SI{532}{nm}$ laser gated by a TTL signal. 
The beam is focused onto the diamond using an oil-immersion objective (Olympus UPlanSApo 60x, NA = 1.35, working distance 300 $\mu m$). 
Fluorescence in the red spectral range is collected through the same objective, passed through a $\SI{50}{\mu m}$ pinhole and a 650 nm long-pass filter, and subsequently detected by an avalanche photodiode (APD) connected to a TimeTagger (Swabian Instruments) for photon counting. 
The diamond is glued to the resonator, and the combined assembly is mounted onto a WR12 rectangular waveguide. 
For nuclear spin control a copper wire is positioned on the backside of the diamond.
The assembly is centered in the room-temperature bore of a superconducting vector magnet (Scientific Magnetics) at \SI{2.7}{T}.

Microwave pulses at $\sim \SI{73}{GHz}$ are generated by mixing a continuous-wave carrier with pulsed modulation. 
Specifically, a carrier frequency (Anritsu MG3697C) is upconverted six times, pre-amplified (S12MS, OML Inc.), and mixed (SAGE-SFB-12-E2) with the pulsed output of an arbitrary waveform generator (AWG, Keysight M9505A; frequency range 100–1000 MHz). 
The resulting microwave signal is further amplified (SAGE-SBP-7137633223-1212-E1 and SAGE-AMP-12-02540) before being coupled to the resonator via the WR12 waveguide. 
RF control of the nitrogen spins is provided by the same AWG and subsequently amplified (AR150A250).
The resonator is fabricated from a double-side polished sapphire substrate ($\alpha$-Al2O3 (0001), 280 µm thick, commercially available). 
The resonator structure is realized by copper plating, photolithography, electroplating, reactive-ion etching, and final laser cutting.
For further details regarding the resonator design, see \cite{Maier.17032025}.


The diamond used in this study was prepared by homoepitaxial growth on a 1.5x1.5 mm$^2$ (111) IIa HPHT substrate (FSBI TISNCM) with a miscut angle of \SI{1.7}{\degree}. 
Before overgrowth, the substrate was chemically cleaned in nitric acid at \SI{250}{\celsius} and subsurface damage from polishing was removed employing an oxygen ICP plasma etch to remove approx. 2-3 $\mu$m of the upmost layer. 
Overgrowth was carried out employing a custom-build diamond CVD reactor with an ellipsoidal resonator for sharp layer interfaces \cite{schatzle_chemical_2023}. 
Low methane process conditions were chosen to enable the growth of aligned NV centers in a step-flow growth mode \cite{miyazaki_atomistic_2014}. 
A layer stack consisting of an approx. \SI{1}{\mu m} thick intrinsic buffer and the nitrogen doped layer with \SI{277}{nm} was grown. 
Both layers employed \SI{0.5}{\percent} CH$_4$ in H$_2$ and growth was carried out at around \SI{850}{\celsius}. 
The N-doped layer was grown using isotopically purified $^{12}$CH$_4$ and nitrogen at an N/C ratio of \SI{40000}{ppm}. 
From secondary ion mass spectrometry (SIMS), a nitrogen concentration of approx. \SI{2.4e18}{N \per cm^3}, i.e., \SI{13.6}{ppm} was obtained. 
This corresponds to a doping efficiency of around \SI{0.03}{\percent} as expected from literature \cite{nakano_impact_2022}.

ODMR characterization of the diamond demonstrated a Rabi contrast of up to \SI{30}{\percent} as expected for preferentially aligned NV centers \cite{michl_perfect_2014}. 
Typical decay times of the NVs are $T_2^* \sim \SI{0.9}{\mu s}$, $T_2^\mathrm{Hahn} \sim \SI{15}{\mu s}$ and $T_1 \sim \SI{3.4}{ms}$. 
The spin-dephasing times obtained at \SI{2.7}{T} correlate to 11 ppm of nitrogen contributing to the spin bath when applying the model described in literature \cite{bauch_decoherence_2020}, which is in good agreement with the measured concentration from SIMS.

\begin{figure}[ht]
    \centering
    \includegraphics{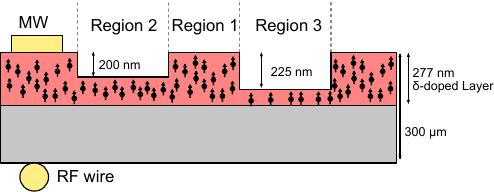}
    \caption{\textbf{Visualization of the diamond sample.}
    The overgrown $\delta$-doped layer contains the $^{14}$NV centers.
    Three different NV concentrations in the confocal spot are prepared by etching of the diamond surface.
    The MW resonator for electron spin manipulation is located on top of the diamond, while the RF wire for nuclear spin manipulation is located at the back.
    }
    \label{fig:Figure_Diamond}
\end{figure}

\section*{Supplementary Note 2: Time averaged spin distribution}
To beat the photon shot noise in the readout of nitrogen spin states $\tilde{J}_z$, the number of repetitive readouts $m$ (or the readout time $T$) has to be increased.
The main limiting factor here, are small disturbances to the $\tilde{J}_z$ induced by the readout, leading to an effective decay constant $T_1$.
Therefore, the readout results in an effective average over the statistical decay process, yielding $\langle \tilde{J}_z\rangle_T$.
The variance of this readout over multiple measurements is given by
\begin{align}
    \text{Var}(\langle \tilde{J}_z \rangle_T) &= \left\langle \langle \tilde{J}_z \rangle_T^2 \right\rangle - \left\langle \langle \tilde{J}_z \rangle_T \right\rangle^2 \\
\end{align}
For now, a single two level $I = 1/2$ system with thermal polarization (i.e. $\left\langle \langle \tilde{J}_z \rangle_T \right\rangle^2 =0$) is assumed.
Thus
\begin{align}
    \text{Var}(\langle \tilde{J}_z \rangle_T) &= \left\langle \left( \frac{1}{T} \int_0^T \tilde{J}_z(t) \mathrm{d}t \right)^2 \right\rangle\\
    &= \left\langle\frac{1}{T^2}\int_0^T \int_0^T \tilde{J}_z(t_1) \tilde{J}_z(t_2) dt_1 dt_2 \right\rangle \\
    &= \frac{1}{T^2}\int_0^T \int_0^T \left\langle \tilde{J}_z(t_1) \tilde{J}_z(t_2) \right\rangle dt_1 dt_2 \\
    &= \frac{2}{T^2}\int_0^T \int_0^{t_2} \left\langle \tilde{J}_z(t_1) \tilde{J}_z(t_2) \right\rangle dt_1 dt_2 \\
    &= \frac{2}{T^2}\int_0^T \int_0^{t_2} \left\langle \tilde{J}_z (t_2-t_1) \tilde{J}_z(0) \right\rangle dt_1 dt_2 \\
    &= \frac{2}{T^2}\int_0^T \int_0^{t_2} \left\langle \tilde{J}_z(\tau) \tilde{J}_z(0) \right\rangle d\tau dt_2 \\
    &= \frac{2}{T^2}\int_0^T \int_\tau^{T} \left\langle \tilde{J}_z(\tau) \tilde{J}_z(0) \right\rangle dt_2 d\tau \\
    &= \frac{2}{T^2}\int_0^T (T - \tau) \left\langle \tilde{J}_z(\tau) \tilde{J}_z(0) \right\rangle d\tau \label{eq:Var_Jz}
\end{align}
The correlation function of the spin state $\tilde{J}_z(t)$ is assumed to be stationary and is given by
\begin{align}
    C(\tau) = \langle \tilde{J}_z(\tau) \tilde{J}_z(0) \rangle = \sigma_{0}^2 e^{-|\tau|/T_1} \label{eq:corr_func}
\end{align}
where $\sigma_{0}^2$ is the base variance of the readout without decay.
Inserting Eq. \ref{eq:corr_func} into Eq. \ref{eq:Var_Jz} yields
\begin{align}
    \text{Var}(\langle \tilde{J}_z \rangle_T) &= \frac{2}{T^2}\int_0^T (T - \tau) \sigma_0^2 e^{-\tau/T_1} d\tau \\
    &= \frac{2\sigma_0^2 T_1^2}{T^2} \left( \frac{T}{T_1} + e^{-T/T_1} -1  \right) \thickspace .
\end{align}
The final expression for the standard deviation of the spin readout $\sigma_{\tilde{J}_z}$ is given as
\begin{align}
    \sigma_{\tilde{J}_z} &= \sqrt{\text{Var}(\langle \tilde{J}_z \rangle_T)} \\
    &= \sigma_0  \sqrt{\frac{2T_1^2}{T^2} \left(\frac{T}{T_1} + e^{-T/T_1} -1\right)} \thickspace .
\end{align}
When reading out the thermal spin state of a spin ensemble, $\sigma_0$ is given by the spin projection noise $\sigma_0 = \frac{1}{N_\mathrm{NV}}\sqrt{N_\mathrm{NV}\frac{I(I+1)}{3}}$.
Experimentally, our spin state readout is limited to the two active NV levels.
Thus, the three spin states of the $^{14}$N nuclear spins ($I = 1$) are mapped onto the electron spin via $|1\rangle_\mathrm{N} \rightarrow |1\rangle_e, |0\rangle_\mathrm{N} \rightarrow |0\rangle_e, |-1\rangle_\mathrm{N} \rightarrow |0\rangle_e$, effectively binning the spin 1 system into a spin 1/2 system.
This leads to an effective narrowing of the observed spin distribution, as well as a shift of the mean value.
This reduces the observed width by an additional factor of $\sqrt{\frac{1}{3}}$, resulting in the final analytical expression
\begin{align}
    \sigma_{\tilde{J}_z} = \sqrt{\frac{I(I+1)}{3}}\sqrt{\frac{1}{N_\mathrm{NV}}} \sqrt{\frac{1}{3}} \sqrt{\frac{2T_1^2}{T^2} \left(\frac{T}{T_1} + e^{-T/T_1} -1\right)}\thickspace . \label{eq:sigma_j_analytical}
\end{align}
This expression captures the spin projection noise (i.e. statistical polarization) of a thermal spin ensemble of spin $I$ particles, as detected in Fig. 2 of the main text.

\begin{figure}[ht]
    \centering
    \includegraphics{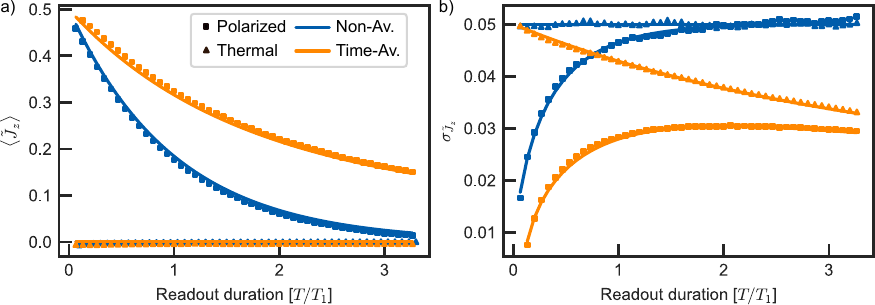}
    \caption{\textbf{Simulation of the effect of $T_1$-decay during the readout on expectation value (panel a) and standard deviation (panel b) of the detected spin distribution.}
    The properties of the spin distribution of $N_\mathrm{NV} = 100$ at time $T$ are visualized in blue, while the distribution obtained by averaging the spin distributions between 0 and $T$ are visualized in orange.
    The properties of a thermal state (triangles) and a polarized state (squares) are shown.
    }
    \label{fig:Figure_T1_effect}
\end{figure}

For completeness, here we also provide the expressions for expectation value and the standard deviation of the non-time averaged distributions $\tilde{J}_z$, as well as the time-averaged distributions $\langle \tilde{J}_z \rangle_T $ with respect to their polarization $p$ in Supplementary Tab. \ref{tab:equations_T1}.
A Monte-Carlo simulation of the ensemble readout of $N_\mathrm{NV} = 100$ spin-1/2 nitrogen spins was performed to visualize the effects of the readout on the underlying spin distributions (Supplementary Fig. \ref{fig:Figure_T1_effect}).
Due to the readout-induced decay of the spin polarization, a clear distinction between the projection noise of thermal and polarized spin distribution is getting increasingly difficult, as the readout time $T$ approaches $T_1$ (orange curves in Supplementary Fig. \ref{fig:Figure_T1_effect} b).
Thus, in all our experiments of the main text the readout duration was fixed to a time significantly shorter than the respective $T_1$.

\begin{table}[]
\centering
\caption{Expectation value and projection noise of the spin distribution during the readout.
Both expressions are given for the spin distribution at time $T$, as well as the time-averaged spin distribution between 0 and $T$.
These equations are used to model the data of Supplementary Fig. \ref{fig:Figure_T1_effect}.}
\label{tab:equations_T1}
\begin{tabular}{c|cc}
    \hline
    Averaging type& Expectation value $\langle \cdots \rangle$ & Projection noise $\sigma$\\ \hline
    Non-averaged & $p I e^{-T/T_1}$& $\sigma_0\sqrt{1 - (p e^{-T/T1})^2}$ \\
    Time-averaged & $p I (1-\frac{T_1}{T} e^{-T/T1})$ & $\sigma_0 \sqrt{\frac{2T_1^2}{T^2}(\frac{T}{T_1} + e^{-T/T1} -1) - (p\frac{T_1}{T}(1-e^{-T/T1}))^2}$\\ \hline
\end{tabular}
\end{table}

\section*{Supplementary Note 3: Spin projection noise transition}
\begin{figure}[ht]
    \centering
    \includegraphics{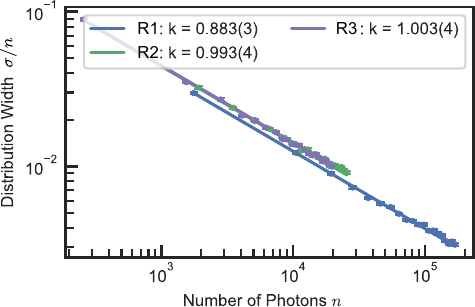}
    \caption{\textbf{Calibration of the photon shot noise.}
    Detected distribution width $\sigma/n$, when subtracting both reference measurements $r_1$ and $r_2$.
    The fit function is $\sigma/n = k\sqrt{2/n}$.
    }
    \label{fig:Figure_k_calibration}
\end{figure}

In the realization of this experiment, the distribution of $\tilde{J}_z$ is optically recorded by detecting the emitted photons of an ensemble of NVs using an avalanche photodiode (APD).
The distribution of detected photons $b - a$ is governed by the variance sum of the photon shot noise $\sigma_n = \sqrt{a+b} =  \sqrt{1-\frac{c}{2}}\sqrt{2n}$ and the spin distribution $\sigma_{\tilde{J}_z}$ via
\begin{align}
    \sigma^2 &= \sigma_n^2 + (2 n c \sigma_{\tilde{J}_z})^2 \thickspace ,
\end{align}
where $c$ is the optical contrast of the NV readout.
Due to the high photon emission rate of the NV ensembles of up to 6000 kilocounts per second, our APD is no longer working in the linear regime, leading to an artificial reduction of the detected photon width according to
\begin{align}
    \sigma^2 &= k^2\left(\sigma_n^2 + (2 n c \sigma_{\tilde{J}_z})^2\right) \thickspace . \label{eq:sigma_final}
\end{align}
The reduction factor $k$ can be easily determined by calibrating the effective photon shot noise in the readout of the reference measurements of the baseline counts (Supplementary Fig \ref{fig:Figure_k_calibration}).
For this, the relative width of the Skellam distribution obtained by subtracting both reference readouts $r_1$ and $r_2$ is determined, and compared to the theoretical photon shot noise of $\sigma_n/n = k \sqrt{2/n}$.
The extracted data show ideal photon shot noise limited behavior as $\sigma_n \propto \sqrt{n}$, showing that our distribution width is not artificially increased, e.g. through slow laser drifts or other noise sources.
Thus, any increase in the detected width of the measurement readout $b-a$ can be attributed to contributions originating from the underlying spin distribution.

\begin{figure}[ht]
    \centering
    \includegraphics{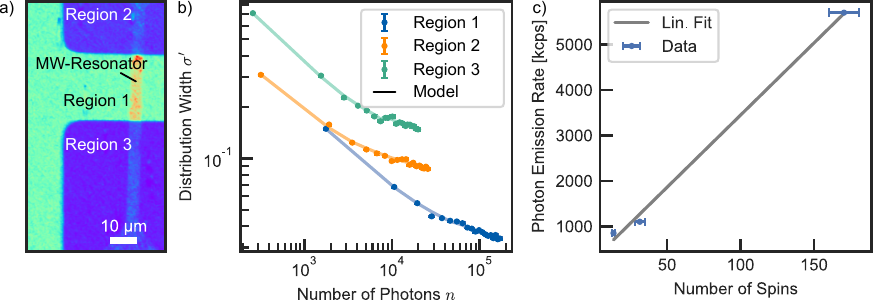}
    \caption{\textbf{Determination of number of NVs in the confocal spot.}
    \textbf{a)} Confocal scan image.
    Three regions with different NV concentrations and the microwave resonator are visible.
    \textbf{b)} transition from the photon shot noise dominated regime to the spin projection noise dominated regime.
    The fit parameters of the model from Eq. \ref{eq:sigma_final} are given in Supplementary Tab. \ref{tab:fit_param}.
    \textbf{c)} Linear relation between the extracted number of NVs and the photon emission rate under laser excitation.
    }
    \label{fig:Figure_S2_n_ssr_sweep}
\end{figure}

The model used to fit the measurement data in Fig. 2 d) in the main text and Supplementary Fig. \ref{fig:Figure_S2_n_ssr_sweep} is given by Eq. \ref{eq:sigma_final} and Eq. \ref{eq:sigma_j_analytical}, with $N_\mathrm{NV}, T_1$ and $k$ as free fit parameters.
The fit results for three regions with different NV concentrations are shown in Supplementary Fig. \ref{fig:Figure_S2_n_ssr_sweep} and the fit parameters are given in Supplementary Tab. \ref{tab:fit_param}.
Instead of the time of the readout $T$ and the time decay $T_1$, the data is fitted as a function of the number of photons $n$ and its decay $n_\mathrm{T_1}$.
The extracted values of the reduction factor $k$ closely match those obtained in the calibration of baseline photon shot noise, further validating the approach.
Over the duration of the measurement, the $T_1$ decay of the nitrogen spin state does not yet significantly affect the observed distribution, leading to a large uncertainty of $T_1$ from the fit.
The extracted number of NVs of 170(10), 31(3) and 14(1) for region 1, 2 and 3, respectively, decrease with decreasing height of the NV layer, as expected.
Typically, optical readout of NV centers is limited by charge state infidelities, as some NV centers are initialized into the optically inactive NV$^0$ charge state, instead of NV$^-$.
It has to be noted that in our approach the NV centers are repeatedly re-initialized over the duration of a single readout loop, so that eventually all NVs contribute to the acquired signal.
As a result this offers a direct method to directly determine the actual number of NV centers in the confocal spot, instead of only the fraction of NV centers in the negative charge state.

Robust validation of our method to determine the total number of NV centers in the detection volume is not trivial, as no other such method exists to this day.
One approach is to compare the extracted numbers with the observed steady state fluorescence emission rate, as more NV centers typically should emit more photons.
However, this approach has to be treated very carefully, as the real emission rate does not necessarily increase linearly with the number NVs in the confocal volume, as the emission rate depends on complicated steady state charge dynamics in the beam profile of the excitation laser.
In our case, the extracted number NVs follows a linear relation to the photon emission rate $r$ under green excitation (Supplementary Fig. \ref{fig:Figure_S2_n_ssr_sweep} c).
The linear fit function is given by $r = n_1^\prime N_\mathrm{NV} + r_0$, where $r_0 = \SI{270(190)}{kcps}$ is the background fluorescence and $n_1^\prime = \SI{32(2)}{kcps/NV}$ is the emission rate per NV.

A second option to roughly estimate the number of NV centers in the detection volume is by geometric considerations.
Assuming a confocal limited spot with a diameter of $d_\mathrm{beam} = \frac{\lambda_0}{0.84 \mathrm{NA}}$ (Gaussian beam profile with $1\sigma$ radius), where $\lambda_0 = \SI{532}{nm}$ is the wavelength of the laser and NA the numerical aperture of the objective, the number of nitrogen atoms $N_\mathrm{N}$ in the \SI{277}{nm} thick NV layer can be estimated by using the nitrogen density $[\mathrm{N}] = \SI{11}{ppm}$ obtained by secondary ion mass spectroscopy (SIMS) and comparison to another calibration sample, yielding $N_\mathrm{N} \approx 36000$.
The rate of NV centers formed for each Nitrogen atom in this non-annealed sample is assumed on the order of \SI{0.3}{\percent} based on previous calibrations, yielding an estimate for the number of NVs in the confocal volume on the order of $N_\mathrm{NV} \approx 140$.
This number is close the extracted number of $N_\mathrm{NV} = \SI{170(10)}{}$, but it is again only a rough estimate as for example the NV conversion rate cannot be determined exactly and additional NVs from outside the $1 \sigma$ region might contribute to the observed signal.
For the thinner, etched regions (Region 2 and 3) this calculation of the average expected number of NV centers cannot be performed, as in these regions inhomogeneous distribution of the NV centers could clearly be seen in the confocal image.

Nevertheless, as both, the comparison to the photon emission rate, and the expected number of NVs from geometric considerations confirm the extracted number of NV centers from the spin projection noise, we are confident in its validity.


\begin{table}[]
\centering
\caption{Fit parameters of projection noise transition fit according to Eq. \ref{eq:sigma_final}}
\label{tab:fit_param}
\begin{tabular}{cccccccc}
    \hline
    Region & Emission Rate [kcps] & $N_\mathrm{NV}$ & $n_{T_1}$ [$10^3$ counts] & $k$ \\ \hline
    1 & 5700 & 170(10) & 582(469) & 0.89(1) \\
    2 & 1100 & 31(3) & 1600(22092) & 0.99(2) \\
    3 & 850 & 14(1) & 3222(2436) & 0.98(1) \\ \hline
\end{tabular}
\end{table}

\section*{Supplementary Note 4: $T_1$-decay of spin polarization}
At a high magnetic field of $\SI{2.7}{T}$ the nitrogen spin states are stabilized to achieve a quantum-non-demolition measurement condition.
However, during the large number of readout repetitions applied in this work small perturbations add up and the spin state shows a characteristic $T_1$-decay, thus setting an upper limit to the possible number of readouts.
During the duration of the readout $T$ the polarization $p$ decays to the steady-state thermal polarization $p_\mathrm{ss}$ according to the exponential decay $p = p_0(1-p_\mathrm{ss}) e^{-T/T_1} + p_{ss}$.
Since our readout bins both, $|0\rangle$ and $|-1\rangle$ into the same electron spin state, the polarization decays to a steady state polarization of $p_\mathrm{ss} = -\frac{1}{3}$.
Averaging of the spin state decay during the full readout duration $T$ yields the observed polarization
\begin{align}
    p_\mathrm{obs} &= \frac{1}{T}\int_0^T p dT = p_0(1-p_\mathrm{ss}) \frac{T_1}{T} \left( 1 - e^{-T/T_1}\right) + p_\mathrm{ss} = p_0(1-p_\mathrm{ss}) \frac{m_{T_1}}{m} \left( 1 - e^{-m/m_{T_1}}\right) + p_\mathrm{ss} \label{eq:T1_decay} \thickspace.
\end{align}
The final expression is expressed in terms of the readout repetition number $m$ instead of the total readout duration $T$.
The polarization decay under readout for the nitrogen spin states $|1\rangle$ and $|0\rangle$ are shown in Supplementary Fig. \ref{fig:Figure_S3_T1_decay}.
The $|0\rangle$ state shows a twice faster decay compared to the $|1\rangle$ state, mainly because of additional relaxation pathways through the spin-allowed transitions into both, $|1\rangle$ and $|-1\rangle$ states.
This faster decay rate in $|0\rangle$ is the main reason for the larger spin projection noise during the inversion of the nitrogen rabi drive in Fig. 3 b) of the main text.
\begin{figure}[ht]
    \centering
    \includegraphics{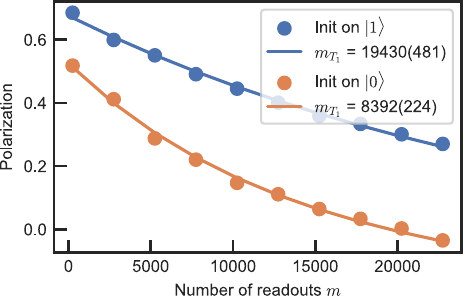}
    \caption{\textbf{Nitrogen Spin Polarization Decay.}
    The nitrogen spin level $|1\rangle$ and $|0\rangle$ show a decay to 1/e after $m_\mathrm{T_1} = 19430$ and $8392$ readouts, respectively.
    The experimental data is fitted by Eq. \ref{eq:T1_decay}.
    }
    \label{fig:Figure_S3_T1_decay}
\end{figure}

\section*{Supplementary Note 5: Noise spectroscopy using dynamical decoupling}
When using noise spectroscopy based on dynamical decoupling (DD), the sensor is selectively coupled to the frequency of an oscillating noise source with the signal
\begin{align}
    S = B_\mathrm{osc}\cos{(2\pi f t + \lambda)} \thickspace ,
\end{align}
where $B_\mathrm{osc}$ is the magnetic field amplitude, $f$ is the frequency and $\lambda$ is the random phase of the signal.
By sweeping the inter-pulse delay $\tau$ of the decoupling sequence, the sensor spin along the +Y axis acquires a phase \cite{vorobyov_transition_2023}
\begin{align}
    \theta = \alpha \mathrm{sinc}((\tau-\tau_0)N_p\tau\pi) \sin(\lambda) = \alpha^\prime \sin(\lambda) \thickspace .
\end{align}
Here $\tau_0 = \frac{1}{2f}$ is resonant pulse spacing, $N_p$ is the number of applied $\pi$ pulses and $\alpha = 2 \pi \frac{2}{\pi} B_\mathrm{osc} \gamma_e \tau_\mathrm{sens}$ is the interaction strength between the electron spin and the oscillating target field during the sensing time $\tau_\mathrm{sens}$ \cite{degen_quantum_2017}. 
$\gamma_e = \SI{28.04}{GHz/T}$ is the gyromagnetic ratio of the electron spin.
The expectation values along X and Y are given by 
\begin{align}
    \langle X \rangle &= \langle \sin{(\theta)}\rangle_\lambda = 0 \thickspace ,\\
    \langle Y \rangle &= \langle \cos{(\theta)}\rangle_\lambda = 0.5 J_0(\alpha^\prime) \thickspace ,
\end{align}
where $\langle \rangle_\lambda$ is the average over the random phase $\lambda$ and $J_0$ is the zeroth order Bessel function of the first kind.
The standard deviation of the X and Y readout are then given by 
\begin{align}
    \sigma_x &=  0.5 \sqrt{\frac{1 - J_0(2\alpha^\prime)}{2}} \thickspace ,\\
    \sigma_y &=  0.5 \sqrt{\frac{1 + J_0(2\alpha^\prime)}{2} - J_0(\alpha^\prime)^2} \thickspace .
\end{align}
The projection of the spin ensemble along Z remains unaffected by the DD-detection sequence in the unpolarized state ($\langle Z \rangle = 0$, $\sigma_z = \sigma_\mathrm{thermal}$).
Under our experimental conditions, the observed standard deviation of the distribution is influenced by the remaining photon shot noise via $\sigma^\prime = \sqrt{\left(\frac{\sigma_n}{2nc}\right)^2 + \sigma_{x/y/z}^2}$.
For fitting the experimental data in Fig. 4 c) in the main text, an additional scaling parameter $k_1$ was introduced to account for readout infidelities leading to a reduction in the amplitudes:
\begin{align}
    \sigma^\prime = \sqrt{\left(\frac{\sigma_n}{2nc}\right)^2 + k_1\sigma_{x/y/z}^2}
\end{align}
In the experiments depicted in Fig. 4 of the main text, a frequency of $f = \SI{250}{kHz}$ and a signal amplitude of $B_\mathrm{osc} = \SI{1.84(7)}{\mu T}$ at the position of the NV center was generated by the AWG and supplied by the same wire as the RF for the nuclear spin drive.
A standard XY8 decoupling sequence was applied to detect the signal.

\section*{Supplementary Note 6: Reconstruction of the Husimi $Q$-function}
In our experiment, the electron spin interacts with coherent RF field with a random phase, leading to a redistribution of the initialized spin state.
To reconstruct the Husimi $Q$-function of the collective spin state, we measure a set of marginal spin distributions $P_S(m_J, \theta, \phi)$ along different quantization axes, defined by the azimuthal and polar angles $\theta$ and $\phi$, as well as the magnetic quantum number $m_J \in [-J .. J]$ along the chosen axis.
Namely, these are one projection along the Z-axis ($\theta = 0^\circ$) and nine equally spaced projection axes in the equatorial plane ($\theta = 90^\circ, \phi \in [-90^\circ, 90^\circ]$).
Each measured distribution corresponds to the projection of the quantum state aligned with the chosen axis.
The obtained histogram data $(b - a)/(2nc)$ is a convolution of the photon shot noise, given by a Skellam distribution, and the underlying spin marginal distribution $P_S$.
For the reconstruction of the spin states, first the deconvolution is achieved using a maixmum-likelihood algorithm.
In the next step, the obtained distribution was normalized by the obtained borders of the nitrogen Rabi measurement (Fig. 3 a) of the main text)relaxometry chauvet, to isolate only active spins (i.e. NV centers with the correct charge and spin state) and remove NV centers where the nitrogen spin was initialized into a state other than $|\uparrow \rangle$, since these spins remain in the polarized state.
From these normalized marginal distributions, we employ an inversion algorithm that maps the experimentally acquired probability distributions onto the Husimi $Q$-function \cite{carmichael_statistical_2008}, defined as
\begin{align}
    Q(\theta, \phi) = \frac{1}{\pi} \langle \theta, \phi| \rho| \theta, \phi \rangle \thickspace ,
\end{align}
where $|\theta, \phi \rangle$ are spin-coherent states on the Bloch sphere \cite{amiet_coherent_1999}.
This procedure yields a quasiprobabilistic distribution that provides a phase-space representation of the measured state, allowing us to directly visualize its coherence properties, quantum fluctuations and nonclassical features.
In practice, the total spin state $\rho$ is expanded as a sum of $i = 51$ spin coherent states $\rho_\mathrm{coh}$ and a thermal part $\rho_\mathrm{therm}$:
\begin{align}
    \rho = \sum_{k=0}^i a_k \rho_\mathrm{coh}(\theta_k, \phi_k) + a_\mathrm{therm}\rho_\mathrm{therm} \thickspace ,
\end{align}
where $a_k, a_\mathrm{therm}$ are the respective weights.
The choice of potential spin coherent states is physically informed and limited to states equally spaced in a polar angle range of $[- \Delta \phi, \Delta \phi]$ around $\phi = 90^\circ$ for a single $\theta$ with equal weights $a_k$, as expected for a spin system interacting with an oscillating field.
A maximum likelihood algorithm solves for the optimum values of $\Delta \phi$, $\theta$ and $a_\mathrm{therm}$ for a given number of spins $N_\mathrm{NV}$ (here $N_\mathrm{NV} = 26$) in the Dicke basis \cite{dickecoherence1954}.
As shown in Supplementary Fig. \ref{fig:Figure_reconstruction}, the reconstructed spin state $\rho$ reproduces the measured marginal distributions well.
From the obtained spin coherent states $\sum_{k=0}^i a_k \rho_\mathrm{coh}(\theta_k, \phi_k)$ the Husimi $Q$-function can be easily calculated and is visualized on a sphere.
The thermal contribution is a constant offset in the Husimi $Q$-function and is therefore omitted in the surface plots.


\begin{figure}[ht]
    \centering
    \includegraphics{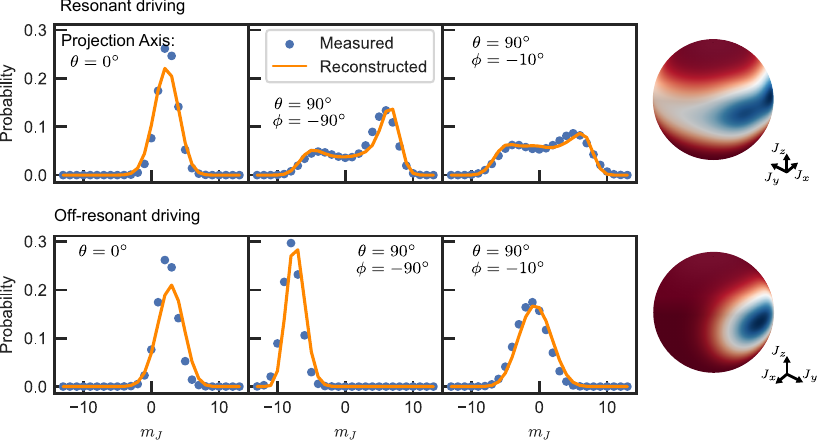}
    \caption{\textbf{Reconstruction of the spin state.}
    Comparison of the measured marginal distributions (blue) and the reconstructed marginal distributions (orange) along three different projection axes.
    The Husimi $Q$-function of the coherent part of the reconstructed spin state is visualized on a sphere.
    }
    \label{fig:Figure_reconstruction}
\end{figure}




\section*{Supplementary Note 7: Sensitivity improvement through projection noise limited readout}
To investigate the applicability of the repetitive readout of ensemble spin systems to sensing protocols, here we calculate the sensitivity and compare it to the conventional single readout.
Generally, the conventional readout scheme is fast but limited to the photon shot noise, while the repetitive readout can reach the spin projection noise at the cost additional experimental overhead for memory spin initialization and increased readout duration.
The sensitivity $\eta$ is defined as the minimal detectable signal $\delta B$ with a signal-to-noise ratio of 1 during the measurement time $T_\mathrm{meas}$ \cite{degen_quantum_2017}
\begin{align}
    \eta &= \delta B \sqrt{T_\mathrm{meas}} =  \frac{1}{\gamma_e c_\mathrm{eff} n \sqrt{\tau_\mathrm{sens}}} \sqrt{\frac{\tau_\mathrm{sens} + \tau_\mathrm{other}}{\tau_\mathrm{sens}}} \sigma\\
    \delta B &= \frac{1}{2\pi\gamma_e c_\mathrm{eff} n \tau_\mathrm{sens}} \sigma \\
    T_\mathrm{meas} &= \tau_\mathrm{sens} + \tau_\mathrm{other}
\end{align}
Here slope detection is assumed with a linear sensor response, and $\sigma$ is the total noise of the readout, $\gamma_e=\SI{28.04}{GHz/T}$ the gyromagnetic ratio of the electron spin, $c_\mathrm{eff}$ the effective contrast, $\tau_\mathrm{sens}$ the sensing time and $\tau_\mathrm{other}$ experimental overhead (e.g. duration of initialization and readout).
In the conventional readout, the noise is given by $\sigma \approx \sqrt{n}$, where $n = n_1$ is the photons detected in a single readout and the effective contrast is equal to the optical contrast of the NV center $c_\mathrm{eff} = c$.
The experimental time overhead $\tau_\mathrm{other}$ consists only of the duration of a single readout $\tau_\mathrm{r}$, leading to
\begin{align}
    \eta_\mathrm{conv} = \frac{1}{2\pi\gamma_e c n_1 \sqrt{\tau_\mathrm{sens}}} \sqrt{\frac{\tau_\mathrm{sens} + \tau_\mathrm{r}}{\tau_\mathrm{sens}}} \sqrt{n_1} \thickspace.
\end{align}

In the repetitive readout, a double-sided scheme is applied (i.e. using the photon count difference $b-a$), doubling the effective contrast $c_\mathrm{eff} = 2c$, while increasing the photon shot noise by a factor of $\sqrt{2}$.
Additionally, the effective contrast is reduced by the decay of the nitrogen spin state during the $m$ readouts with the characteristic decay rate of $m_{T_1}$, leading to $c_\mathrm{eff} = 2c \frac{m}{m_{T_1}} \left(1 - e^{-m/m_{T_1}} \right)$.
The noise $\sigma$ in the readout of a superposition spin state (i.e. maximum spin projection noise) of a spin 1/2 system is given by 
\begin{align}
    \sigma &= \sqrt{\sigma_\mathrm{PSN}^2 + (c_\mathrm{eff}n \sigma_{\tilde{J}_z})^2} \\
    &= \sqrt{ \left( \sqrt{2 n_1 m}\right)^2 + \left( c_\mathrm{eff}n_1 m \sqrt{\frac{1}{2 N_\mathrm{NV}}}\right)^2} \thickspace.
\end{align}
The experimental overhead is given by the time needed for the initialization of the nitrogen spin states ($\tau_\mathrm{init}$) and the nitrogen RF gate to map the electron spin onto the nitrogen spin ($\tau_\mathrm{RF}$) and the duration of the readout ($\tau_\mathrm{r} = m \tau_\mathrm{r}^\mathrm{rep}$) leading to a final expression of 
\begin{align}
    \eta_\mathrm{rep} = &\frac{1}{4\pi\gamma_e c \frac{m}{m_{T_1}} \left(1 - e^{-m/m_{T_1}} \right) n_1 \sqrt{\tau_\mathrm{sens}}} \sqrt{\frac{\tau_\mathrm{sens} + \tau_\mathrm{init} + \tau_\mathrm{RF} + m \tau_\mathrm{r}^\mathrm{rep}}{\tau_\mathrm{sens}}} \\
    &\sqrt{ \left( \sqrt{2 n_1 m}\right)^2 + \left( 2c \frac{m}{m_{T_1}} \left(1 - e^{-m/m_{T_1}} \right)n_1 m \sqrt{\frac{1}{2 N_\mathrm{NV}}}\right)^2}
\end{align}
Since the sensitivity of the spin state readout using the repetitive readout scheme is limited by the spin projection noise, one possibility to improve it is to reduce the projection noise through spin squeezing.
The squeezing parameter $\xi^2$ reduces the spin projection noise by $10^{-\xi}$, while increasing the experimental overhead by the time $\tau_\mathrm{sq}$ to prepare the squeezed state.
In this work, parameters for $\xi^2$ and $\tau_\mathrm{sq}$ were taken from \cite{wu2025spin}
A summary of the parameters used for the sensitivity calculation for Fig. 5 of the main text are given in Supplementary Tab. \ref{tab:sensitivity_param}.

\begin{table}[]
\centering
\caption{Parameters for sensitivity calculations.}
\label{tab:sensitivity_param}
\begin{tabular}{cccccccc}
    \hline
    $\tau_r$ [$\mu$s] & $\tau_r^\mathrm{rep}$ [$\mu$s] & $\tau_\mathrm{init}$ [$\mu$s] & $\tau_\mathrm{RF}$ [$\mu$s]& $\tau_\mathrm{sq}$ [$\mu$s] & $n_1$ [kcps] &$c$ [\%] & $m_{T_1}$\\ \hline
    1.0 & 7.5 &5236 & 600 & 3.0 & 0.036 $N_\mathrm{NV}$ & 15& 50000\\ \hline
\end{tabular}
\end{table}


\newpage
\bibliographystyle{apsrev4-2}
\bibliography{references.bib}